\documentclass[12pt]{iopart}
\usepackage{graphicx,epsf}
\begin{document}

\title[A 128 pixel liquid linear array for radiotherapy quality assurance]{Development and operation of a pixel segmented liquid-filled linear array for radiotherapy quality assurance}

\author{J Pardo\dag,  L Franco\dag, F G\'{o}mez\dag, A Iglesias\dag, A Pazos\dag, J Pena\dag, R Lobato\ddag, J Mosquera\ddag,
  M Pombar\ddag  and J Send\'{o}n\ddag}

\address{\dag\ Departamento de F\'{\i}sica de Part\'{\i}culas, Facultade de F\'{\i}sica, 15782 Santiago de Compostela, Spain}
\address{\ddag\ Hospital Cl\'{\i}nico Universitario de Santiago}
\ead{juanpm@usc.es (J Pardo)}

\begin{abstract}
A liquid isooctane (C$_{8}$H$_{18}$) filled ionization linear array for radiotherapy quality assurance has been designed, built and tested. The
 detector consists of 128 pixels, each of them with an area of 1.7 mm $\times$ 1.7 mm and a gap of 0.5 mm. The small pixel size makes the detector
 ideal for high gradient beam profiles like those present in Intensity Modulated Radiation Therapy (IMRT) and radiosurgery. As read-out electronics
 we use the X-Ray Data Acquisition System (XDAS) with the Xchip developed by the CCLRC.

Studies concerning the collection efficiency dependence on the polarization voltage and on the dose rate have been made in order to optimize the device operation.
 In the first tests we have studied dose rate and energy dependences, and signal reproducibility. Dose rate dependence was found lower than 2.5 \% up to
 5 Gy min$^{-1}$, and energy dependence lower than 2.1 \% up to 20 cm depth in solid water. Output factors and penumbras for several rectangular fields have been
 measured with the linear array and were compared with the results obtained with a 0.125 cm$^{3}$ air ionization chamber and radiographic film, respectively. Finally,
 we have acquired profiles for an IMRT field and for a virtual wedge. These profiles have also been compared with radiographic film measurements. All the comparisons
 show a good correspondence. Signal reproducibility was within a 2\% during the test period (around three months).
 The device has proved its capability to verify on-line therapy beams with good spatial resolution and signal to noise ratio.

\end{abstract}
\submitto{PMB}

\maketitle

\section{Introduction}

Nowadays the verification of radiotherapy treatments in most of the hospitals is performed through air or solid state ionization
 chambers. These chambers are mechanically displaced to obtain beam profiles. IMRT techniques require detectors able to verify
 and to monitor the clinical beams with high spatial resolution and fast response. Furthermore, the dose rate at any point must
 be integrated over the entire exposure, limiting the use of typical ionization chambers. IMRT verification with radiographic
 films (RGFs), radio-chromic films (RCFs)  or electronic portal imaging devices (EPIDs) provide a high spatial resolution. However,
 RGFs need a chemical processing and over-respond to low energy scattered photons (Sykes \etal 1999, Martens \etal 2002), RCFs
 present response non-uniformity (Niroomand-Rad \etal 1998) and calibration of EPIDs is a difficult task, which complicates high precision dosimetry
 with all of these devices. Segmented anode ionization chambers, like those presented in Martens \etal (2001), Belletti \etal (2001)
 and Eberle \etal (2003), and diode arrays (Jursinic and Nelms 2003) are an alternative. Although faster verification procedures
 are possible with these devices, none of them achieve a millimeter range spatial resolution.

In this paper we present the design, the operation principles and the first tests of a 128 pixel linear array whose aim is to obtain
 a profile in a single beam shot with enough resolution to make mechanical displacement unnecessary. Each pixel has an area of 1.7 mm
 $\times$ 1.7 mm. The active medium is a 0.5 mm thick isooctane layer, which is encapsulated between two printed circuit boards. We
 used a standard  liquid isooctane from Merk\footnote{Merk Uvasol quality grade isooctane}, with a purity $\geq$ 99.8\%. No further
 purification, in order to obtain an ultra-pure liquid, has been made. Non polar liquids are becoming and alternative to air and solid
 state semiconductors in radiotherapy detectors due to their tissue equivalent behavior, their sensitivity and their small directional
 dependence. Liquid filled ionization chambers are currently used in radiotherapy both for dosimetry, as shown by Wickman and Nystr\"{o}m
 (1992) and Wickman \etal (1998), and portal imaging as in the device of van Herk and Meertens (1988). One of the most commonly used 
liquids is isooctane (2,2,4 trimethylpentane). This non-polar liquid has a quite constant stopping power ratio to water in a very wide
 energy spectrum (less than 3\% variation from 0.1 MeV to 20 MeV) and also its intrinsic mass density allows to achieve a spatial
 resolution in the millimeter range for therapy beams.

\section{Detector description}
\subsection{Detector design}

\begin{figure}
\begin{center}
\includegraphics*[width=7.5cm]{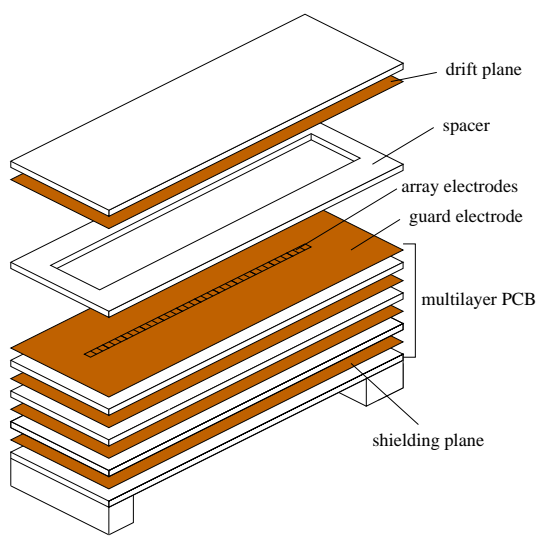}
\end{center}
\caption{Detector scheme. It shows the top PCB, the PEEK spacer and the four layer bottom PCB. }
\label{fig1}
\end{figure}

\begin{figure}
\begin{center}
\includegraphics*[width=8.5cm]{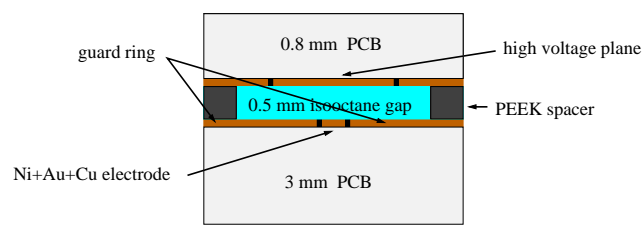}
\end{center}
\caption{Scheme of the detector cross section.}
\label{fig2}
\end{figure}

The linear array has been constructed using two printed circuit boards (PCB) that surround a 0.5 mm thick isooctane layer. The isooctane
 gap is provided by a PEEK\footnote{Poly Ether Ether Ketone} spacer. The chamber walls were fabricated using FR4 fiber glass epoxy. The
 upstream wall has a 0.8 mm thickness and contains the high voltage plane. The downstream one is a four layer PCB with a 3 mm thickness.
 The top layer contains the Cu+Ni+Au anode segmented in 128 pixels. Each electrode has an area of 1.5 mm $\times$ 1.5 mm and is surrounded
 by a guard electrode biased to +2 V. The pitch is 1.7 mm,  and so the linear array consists of 128 cells of 1.7 mm $\times$ 1.7 mm
 $\times$ 0.5 mm giving a total active length of 21.6 cm.  The internal layers contain metallic strips that carry out the ionizing
 charge produced in the liquid to one of the device sides, where the detector is connected to the read-out electronics.  In the bottom
 layer was deposited a 35 $\mu$m thick Cu clad to shield the strips from external noise. The high voltage electrode dimensions (250 mm
 $\times$ 15 mm) are larger than the sensitive area  in order to guarantee a high electric field uniformity in the active volume.
 Figure \ref{fig1} shows a scheme of the detector layout, and figure \ref{fig2} shows the detector cross section. The total dimensions
 of the assembled device are 350 mm $\times$ 70 mm $\times$ 4.5 mm.

\subsection{Read-out electronic system}
\begin{figure}
\begin{center}
\includegraphics*[width=8cm]{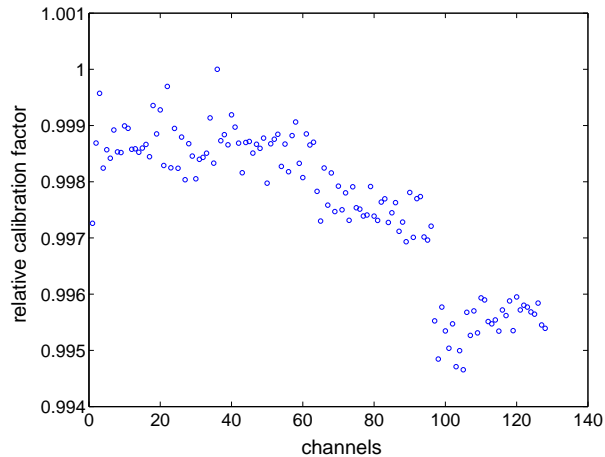}
\end{center}
\caption{Calibration of the XDAS board. The relative non-uniformity is lower than 0.6 \%.}
\label{fig3}
\end{figure}

The X-ray Data Acquisition System (XDAS) has been used as read-out electronic. This system is provided by the company Electron Tubes Ltd.,
 and it is based on the Xchip developed by the CCLRC. It consists of a modular system in which each board has 128 read-out channels, and
 up to 63 boards can be serially connected, giving a maximum of 8064 readout channels. The main characteristics of the XDAS system are
 showed in table \ref{table1}. For this application we only use one board (128 channels). The response of each read-out channel has been
 studied using a Thevenin current source. The mean sensitivity of the channels is 4272$\pm$6 ADC counts per pC. The relative non-uniformity
 in the response of the channels (figure \ref{fig3}) is lower than 0.6 \%.

\begin{figure}
\begin{center}
\includegraphics*[width=7cm]{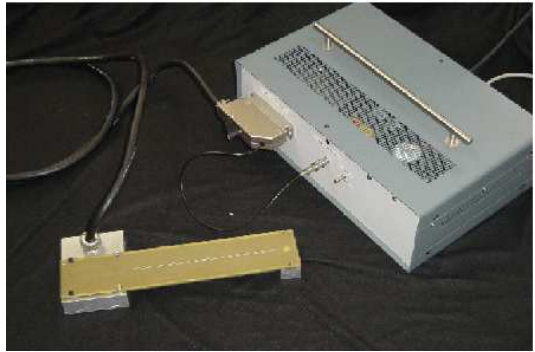}
\end{center}
\caption{Photo of the assembled device. It shows the detector, the cable and the electro-meters station.}
\label{fig4}
\end{figure}

 The XDAS system together with the DC power supplies and a high voltage generator were mounted into a metallic box (the electro-meters
 station) to protect them from external noise and also to make a manageable device. This portable unit is placed close to the detector
 and outside of the direct beam. It is connected to the detector through a 3 meter double shielded cable, and to a PC standard serial
 and parallel ports for digital control and read-out. Figure \ref{fig4} shows a photo of the assembled device.

\begin{table}
\caption{\label{table1}XDAS main characteristics.}
\begin{indented}
\item[]\begin{tabular}{@{}llll}
\br
integration period& 0.01 ms to 0.5 s\\
sub-samples& 256 max.\\
signal to noise ratio& 30000:1\\
readout rate & 5 MB/s max.\\
non-linearity & $<$ 0.1 \% \\
A/D conversion & 14 bit\\
data output & 16 bit\\
dimensions & 101 mm $\times$ 164 mm\\
\br
\end{tabular}
\end{indented}
\end{table}

\section{Principles of operation}
\subsection{Initial recombination}
When ionizing radiation interacts with a medium ionizes electron-ion pairs along its track. Electrons released from molecules thermalize
 at a distance $r$, where the electron and the positive ion are still bounded by the Coulomb interaction. This causes the recombination
 of a fraction of the primary ionization pairs produced, which is called initial recombination. These effects are much more relevant in
 liquids than in gases due to the fact that mass density of liquid hydrocarbons is almost three orders of magnitude higher than density
 of gases at normal conditions, and then $r$ is much smaller. The amount of electron-ion pairs escaping initial recombination per 100 eV
 of absorbed energy is denominated the free ion yield and is denoted as $G_{\rm{fi}}$. The initial recombination, and thus the $G_{\rm{fi}}$,
 depends on the liquid properties, on its temperature, $T$,  an on the external electric field, $E$, (Onsager 1938), but does not on the
 dose rate. For low electric field values ($E\sim$ $10^{3}$ V mm$^{-1}$) the $G_{\rm{fi}}$ rises approximately linear with the electric field:

\begin{equation}
G_{\rm{fi}}\simeq G_{\rm{fi}}^{\rm{0}} \; \lbrack1+aE\rbrack
\label{Gfi}
\end{equation}
The constant $a$ must be measured for a correct absolute dosimetry, but is well approximated by $a\simeq 1/E_{\rm{0}}$ (Mozumder 1974,
 Pardo \etal 2004), where $E_{\rm{0}}=8\pi\epsilon_{\rm{r}}\epsilon_{\rm{0}}(\kappa T)^{2}/e^{3}$ is the called Onsager field. Here
 $\epsilon_{\rm{r}}$ and $\epsilon_{\rm{0}}$ are the relative dielectric constant of the liquid and the dielectric constant of the vacuum
 respectively, $\kappa$ is the Boltzmann constant and $e$ is the electron charge.

\subsection{Volume recombination and ion collection efficiency}
\begin{table}
\caption{\label{table2}Measured charge carriers mobilities ($k_{\rm{-}}$, $k_{\rm{+}}$), volume recombination
 constant ($\alpha$), free ion yield at zero electric field ($G_{\rm{fi}}^{\rm{0}}$) for non ultra-pure isooctane,
 and its relative dielectric constant ($\epsilon_{\rm{r}}$) and Onsager field ($E_{{\rm{0}}}$).}
\begin{indented}
\item[]\begin{tabular}{@{}llll}
\br
$k_{\rm{-}}$ (m$^{3}$ s$^{-1}$ V$^{-1}$)$^{(a)}$ & 3.5$\times$10$^{-8}$ \\
$k_{\rm{+}}$ (m$^{3}$ s$^{-1}$ V$^{-1}$)$^{(a)}$ & 2.3$\times$10$^{-8}$ \\
$\alpha$ (m$^{3}$ s$^{-1}$)$^{(a)}$ & 5.4$\times$10$^{-16}$ \\
$G_{\rm{fi}}^{\rm{0}}$ $^{(b)}$ & 0.32 (20 $^{0}$C)\\
$\epsilon_{\rm{r}}$ & 1.94 (20 $^{0}$C)\\
$E_{{\rm{0}}}$ (V mm$^{-1}$) & 1.74$\times$10$^{3}$ (20 $^{0}$C)\\
\br
$^{(a)}$Determined measuring the temporal development of the read-out\\
 signal in a pulsed beam. The mobilities reported by several authors\\
 for non ultra-pure isooctane are in the range 1-4$\times$10$^{-8}$ m$^{3}$ s$^{-1}$ V$^{-1}$,\\
 probably due to different contamination in the liquids.\\
$^{(b)}$From Pardo \etal (2004).\\
\end{tabular}
\end{indented}
\label{table}
\end{table}
The electrons that have escaped from initial recombination flow due to drift and diffusion, and this made possible the interaction
 between ions from different tracks, which causes the volume recombination. Volume recombination depends on the liquid properties,
 on the electric field, on the dose rate and also on the form in which the dose is delivered (i.e. pulsed or continuous radiation).

Actual clinical linear accelerators delivered the dose in high ionization pulses of a few $\mu$s duration and several ms period.
 The beam dose rate is modulated varying the pulse period.

 If the pulse period, $p$, is higher than the charge carriers drift time in the liquid, i.e. when

\begin{equation}
p\geq \frac{h^{2}}{V k_{\rm{min}}}
\label{p_t}
\end{equation} 
the collection efficiency will not depend on the period (i.e. on the dose rate). In equation (\ref{p_t}) $k_{\rm{min}}$ denotes the lower
 mobility. In this case we can apply the theory of Boag (1950). This theory assumes the negative charge is carried by ions and also neglects
 space charge effects and recombination during the pulse. The theory has been experimentally tested by several authors (see for example
 Johansson \etal 1997), and within it the collection efficiency is given by,

\begin{equation}
f=\frac{1}{u}\ln(1+u)
\label{eq_Boag}
\end{equation} 
with
\[u=\mu\frac{r}{V}h^{2}\]
\[\mu=\frac{\alpha}{e(k_{+}+k_{-})}\]
where $r$ is the amount of charge released by the radiation in the liquid and escaping initial recombination per unit volume and pulse,
 $h$ is the liquid layer thickness, $V$ is the polarization voltage, $k_{+}$ and $k_{-}$ are the mobilities of positive and negative charge
 carriers and $\alpha$ is the volume recombination constant, which for a low permittivity non-polar liquid can be expressed as (Debye 1942),
\begin{equation}
\alpha=\frac{e(k_{+}+k_{-})}{\epsilon_{\rm{r}}\epsilon_{\rm{0}}}
\label{alpha}
\end{equation}

 We use a numerical simulation to calculate the collection efficiency of the detector irradiated by a pulsed beam using the parameters of 
table \ref{table}. We considered the pulse period of a Siemens Primus accelerator placed in the Hospital Clinico Universitario de Santiago,
 which is related to the monitor unit rate as 

\begin{equation*}
\label{eq_period}
p=\cases {(1.93\pm 0.02) \cdot \dot{M}^{-1} & for the 15 MV photon beam\\
(1.08\pm 0.02) \cdot \dot{M}^{-1}  & for the 6 MV photon beam\\}
\end{equation*}
where $p$ is expressed in seconds and $\dot{M}$ is the monitor unit rate expressed in MU min$^{-1}$.

\begin{figure}
\begin{center}
\includegraphics*[height=8cm]{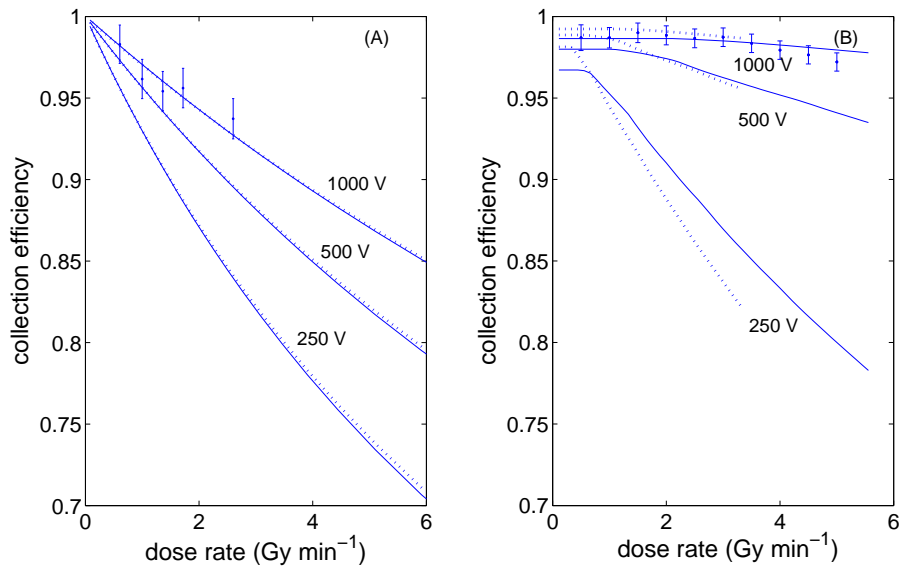}
\end{center}
\caption{(A) Simulated collection efficiency (solid line) plotted against dose rate for several detector operation voltages. The dose rate is modulated
 varying the SDD, and the pulse period is high enough to verify equation (\ref{p_t}) and then the Boag theory (dotted line) can be applied. Experimental
 points for a 1000 V operation voltage are plotted.

(B) Simulated collection efficiency plotted against dose rate for several detector operation voltages with the detector irradiated by 15 MV (solid line)
 and 6 MV (dashed line) beams. The SSD is constant and the dose rate is modulated varying the accelerator monitor
 unit rate. Experimental points for the 15 MV, 1000 V operation voltage are plotted.}
\label{fig5}
\end{figure}

 Figure \ref{fig5} shows the computed detector collection efficiencies. In figure \ref{fig5}(a) the dose rate
 is modulated varying the source-detector distance (SDD). The pulse period verify equation (\ref{p_t}) and then the Boag theory can be applied.
 There is a good correspondence between the simulation, the Boag theory and the experimental points (for a 1000 V operation voltage).
In figure \ref{fig5}(b) the distance is constant and the dose rate is modulated changing the accelerator
 monitor unit rate (up to 300 (500) MU min$^{-1}$ for the 6 (15) MV beam). We can see that in the upper part of the curves there is not
 dependence on the dose rate. This is due to in this region the pulse period is higher than the charge carriers drift time. Due to this
 fact, the collection efficiency does not go to 1 when the dose rate goes to 0, because the zero dose rate limit is achieved taking
 $p\rightarrow\infty$. In the constant part of the curves the results obtained with the simulation are very close to those computed
 from the Boag theory.
 For example, for the 15 MV beam, with a 1000 V polarization voltage, the simulated collection efficiency at low dose rates is $\simeq98.6$
 \%, very close to the experimentally measured ($\simeq98.7$ \%, although due to accelerator dose rate oscillations the experimental results
 have important uncertainties as we can see in figure \ref{fig5}(b)) and to the computed from the Boag theory ($\simeq$ 98.7\%). For the 6
 MV beam with the same voltage the simulated value is $\simeq99.2$ \% and the the computed from the Boag theory is $\simeq99.3$ \%. Only in
 the case of overlapping of charges ionized by different pulses will the collection efficiency depend on the dose rate. In this case the
 Boag theory cannot be applied. In figure \ref{fig5}(b) we can see a quite good agreement between the experimental points, for the 15 MV
 beam and with a 1000 V detector polarization voltage, and the simulation.
 
From figure \ref{fig5} we can conclude that the recombination is higher when there is not pulse overlapping and the dose rate is modulated changing
 the SSD. The maximum detector response non-linearity depends on the charge collection efficiency variation between low and high dose rates.
 Thus, a higher non-linearity is expected when the dose rate is modulated in this way.

\section{First tests of the device}
\subsection{Experimental set-up}
The first tests of the device were performed using a Siemens Primus accelerator placed in the Hospital Cl\'{\i}nico Universitario de
 Santiago. For this accelerator a MU is defined as 1 cGy at the maximum depth (1.6 cm for 6 MV and 3 cm for 15
 MV) in a water phantom for a 10 cm $\times$ 10 cm field and SSD=100 cm. Measurements were performed in a home-made solid water phantom.
 The detector operation voltage was 1000 V and the XDAS integration time was 10 ms. Unless mentioned otherwise, the SSD was 100 cm.

Comparative measurements of OFs and energy dependence have been made with a 0.125 cm$^{3}$ air ionization chamber (PTW, Freiburg,
 Germany, type 31010). Penumbras and profiles measurements were compared with RGFs measurements. In some cases we used a 0.015
 cm$^{3}$ PinPoint chamber (PTW, Freiburg, Germany, type 31006) which has a 2 mm diameter.

\subsection{Pixel response calibration}
To study the pixel response homogeneity the detector was inserted in the phantom at a 2 cm depth and irradiated with 10 cm
 $\times$ 10 cm 6 MV photon shots, each one delivering 50 MU at a 100 MU min$^{-1}$ rate. Between each shot the detector was displaced
 1.7 mm with a micro-metric linear stage in order to compare the read-out signal of each pixel in the center of the field. The maximum
 relative deviation in the response was $\sim$ 6 \%. The non-homogeneity is due to different response of each XDAS read-out channel (studied
 in subsection 2.2.) and to small inhomogeneities in the gap and the pixel area. These effects have been corrected in all the following
 measurements.

\subsection{Read-out signal linearity with the dose rate}
\begin{figure}
\begin{center}
\includegraphics*[width=12cm]{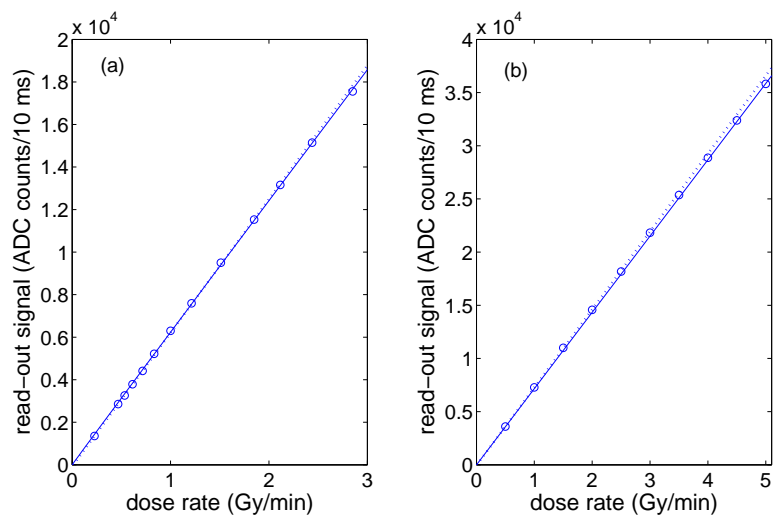}
\end{center}
\caption{Read-out signal plotted versus the dose rate. The dose rate is modulated: (A) varying the SSD for a constant MU rate (100 MU min$^{-1}$); 
(B) varying the MU rate for a constant SSD (100 cm). In both cases the solid line corresponds to the Fowler-Attix fit, and the dotted line to the linear fit 
of the lower dose rate points.}
\label{fig6}
\end{figure}

Figure \ref{fig6} shows the read-out signal in the central pixel of the device plotted against the dose rate, when the detector was
 irradiated by a 10 cm $\times$ 10 cm 15 MV photon beam. The detector is placed in the phantom at a 4 cm depth. In the first case,
 figure \ref{fig6}(a), the monitor unit rate was 100 MU min$^{-1}$ to avoid the superposition of charge carriers ionized by different
 pulses, and the dose rate was modified changing the SSD from 130 cm to 60 cm. In the second case, figure \ref{fig6}(b), the SSD was
 100 cm and the dose rate was modified varying the accelerator MU rate from 50 MU min$^{-1}$ to 500 MU min$^{-1}$ in 50 MU min$^{-1}$ steps. 

It is common to fit this relationship to the empirical expression of Fowler and Attix (1966)
\begin{equation}
S=k\dot{D}^{\Delta}
\end{equation}
where $S$ is the read-out signal, $k$ is a parameter for the detector sensitivity and $\Delta$ a parameter related to the non-linearity
 of the detector response. We obtain $\Delta=0.993\pm0.007$ in the first case, and $\Delta=0.984\pm0.007$ in the second case, which implies
 a small non-linearity in both cases. The linear fit of the lower dose rate points shows 1.5 \% deviation from linearity in the first case (at
2.9 Gy min$^{-1}$) and 2.1 \% in the second (at 5 Gy min$^{-1}$).

\subsection{Photon beams profiles}
\begin{figure}
\begin{center}
\includegraphics*[width=10cm]{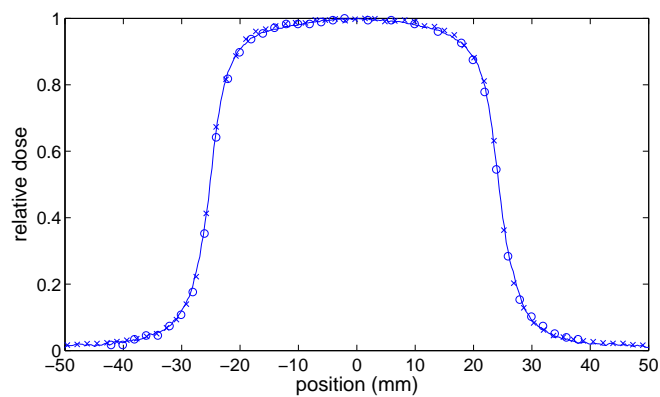}
\end{center}
\caption{Relative dose profile of a 5 cm $\times$ 5 cm 15 MV photon beam at a 5 cm depth in a solid water phantom, measured with the linear array ($\times$), with the PinPoint chamber ($\circ$), with RGF (solid line).}
\label{fig7}
\end{figure}

 Figure \ref{fig7} shows a profile of a 15 MV 5 cm $\times$ 5 cm at 5 cm depth in solid water, measured with our linear array,
 with the PinPoint chamber (displaced with the linear stage in 2 mm steps) and with RGF. All the profiles show a good correspondence.

\begin{figure}
\begin{center}
\includegraphics*[width=12cm]{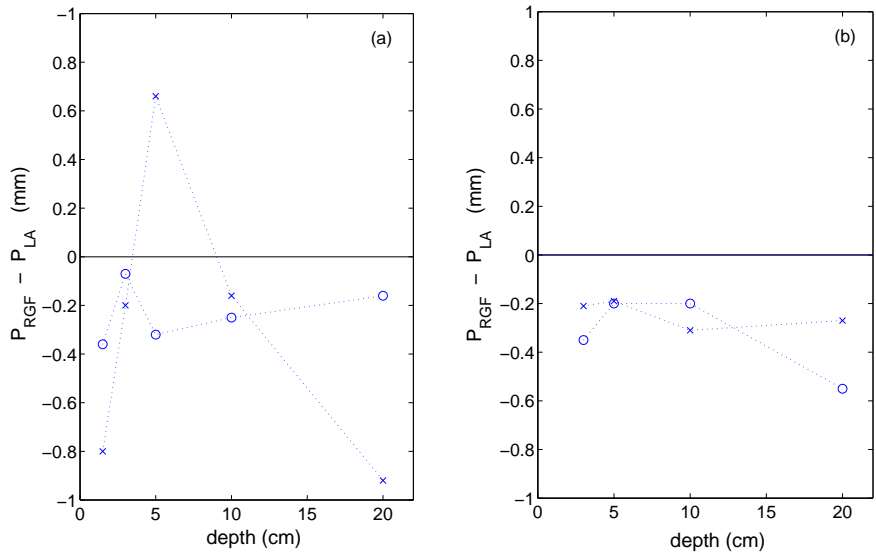}
\end{center}
\caption{Difference between 90\%-10\% ($\circ$) and 80\%-20\% ($\times$) penumbras measured with RGF and with the linear array for 6 MV (a) and 15 MV (b).}
\label{fig8}
\end{figure}

 To study possible systematic deviations in the penumbras measured with the linear array, several 90 \%-10 \% and 80 \%-20 \% penumbras
 of photon beams were measured. The studied fields were 5 cm $\times$ 5 cm at 3 cm, 5 cm, 10 cm and 20 cm for
 15 MV, and at 1.5 cm, 3 cm, 5 cm, 10 cm and 20 cm for 6 MV in the solid water phantom. The MU rate was 100 MU min$^{-1}$ in all cases.
 For each configuration penumbras from linear array profiles were determined through quadratic interpolation, and the average of
 the left and the right penumbras was considered. The results were compared with RGF measurements. We use RGF despite its energy dependence
 because this effects does not affect too much the penumbra measurements, at least at moderate depths and field sizes (Martens \etal 2002), and 
 it provides a high spatial resolution. Differences between measurements of both detectors are plotted in figure \ref{fig8}. We can see that penumbras
 measured with the linear array are broader, in general, than those measured with RGF. Typical uncertainties of the points plotted in figure \ref{fig8} are around $\pm$0.2 mm
 and $\pm$0.4 mm for 80 \%-20 \% and 90 \%-10 \% respectively, and then most of these differences are compatible with zero.

\subsection{Output factors}
Output factors (OFs) are defined as the ratio of the dose at a given depth for a given field size to the dose at the same depth
 for the reference field size. OFs for several rectangular fields were measured with the central pixel of the linear array and
 compared with the OFs measured with the reference detector. The length of the field was 10 cm and the width was varied between
 10 cm and 1 cm in 1 cm steps. A 5 cm $\times$ 5 cm was taken as reference field. The accelerator MU rate was 100 MU min$^{-1}$.
 The depth was 5 cm for 6 MV and 10 cm for 15 MV in the solid water phantom. As reference detector we used the 0.125 cm$^{3}$
 chamber, except for the narrowest field where we used the 0.015 cm$^{3}$ chamber due to its smaller sensitive volume.

\begin{figure}
\begin{center}
\includegraphics*[width=12cm,height=6cm]{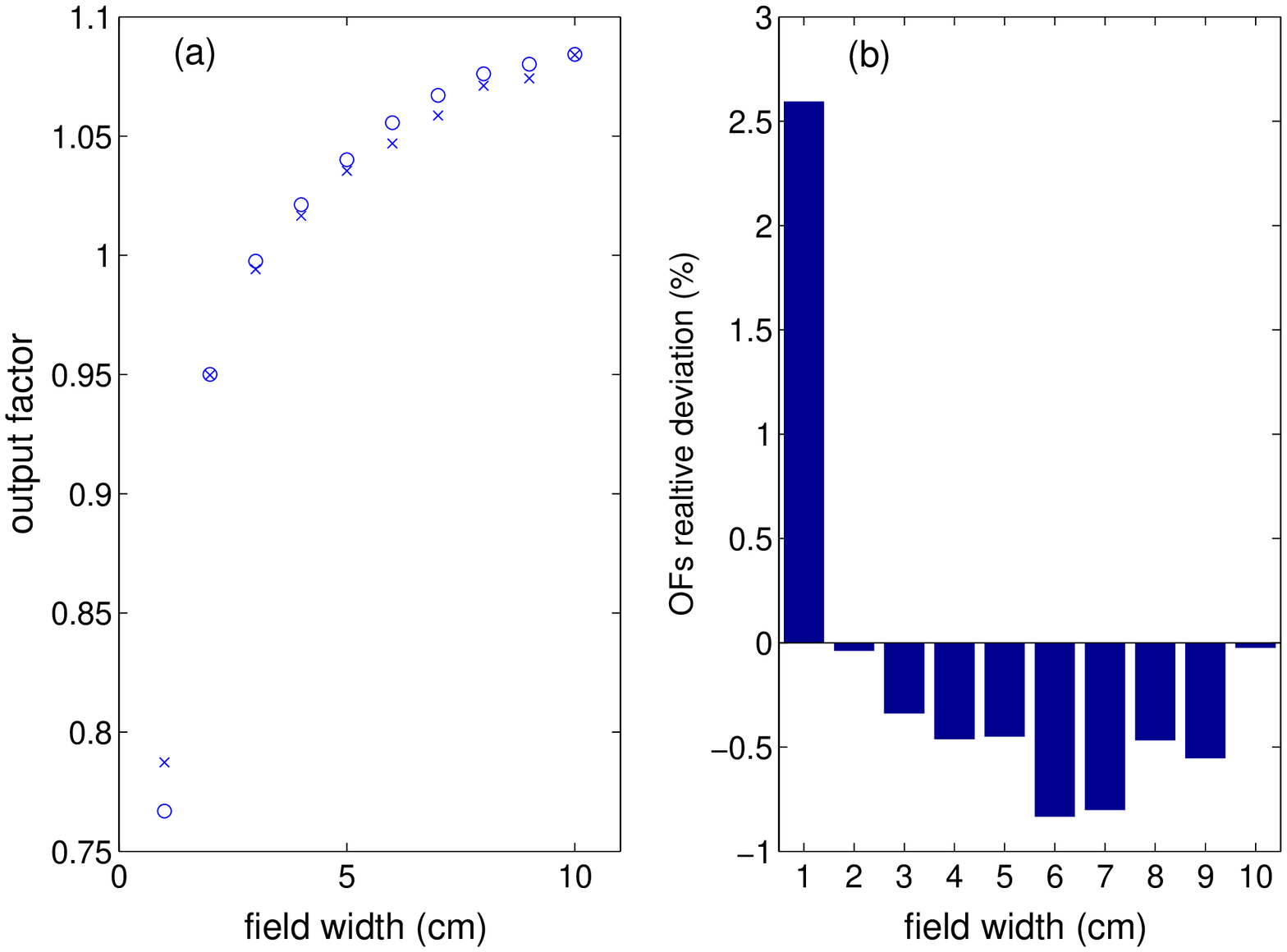}
\includegraphics*[width=12cm,height=6cm]{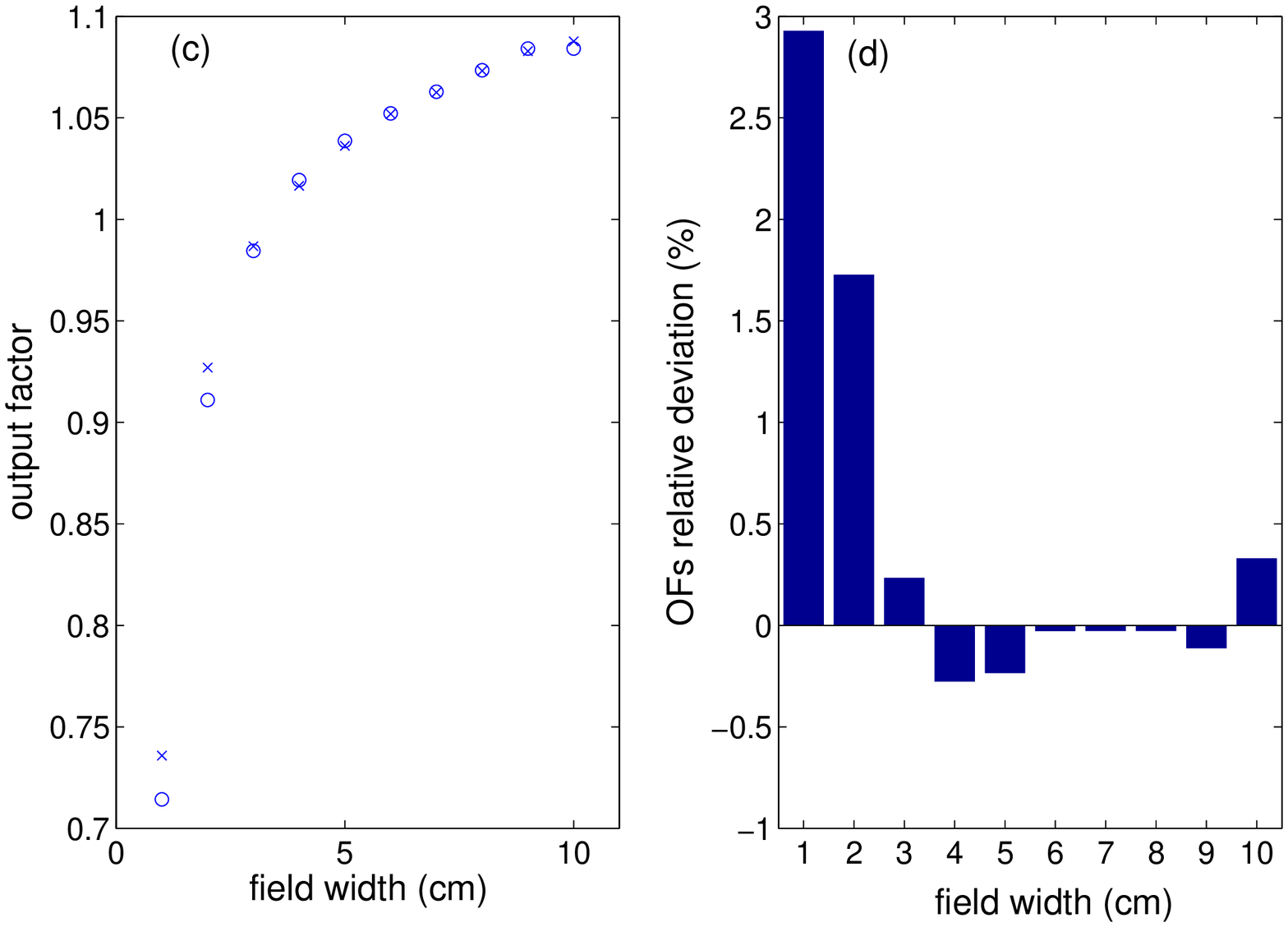}
\end{center}
\caption{OFs measured with the linear array ($\times$) and with the reference detector ($\circ$) for 6 MV (a) and 15 MV (c). OFs relative deviations for 6 MV (b) and 15 MV (d)}
\label{fig9}
\end{figure}

 Figure \ref{fig9} shows the OFs measured with the linear array, the reference detector, and their relative deviation. The
 linear array seems to over-respond to narrow fields both for 6 MV and 15 MV (up to 2.9 \% for the 1 cm width 15 MV field).
 However for the smaller fields the OFs uncertainty is large due to the positioning uncertainty and the difference between the
 sensitive volume of both detectors. For wider fields the linear array presents an under-response (around 0.5 \% and 0.1 \% for 6 MV and 15 MV respectively).
 This behavior has been observed and studied in a similar detector by Martens \etal (2001) who found that is related with the
 effect of the electrode metalization.

\subsection{Energy dependence and effect of the field size.}
\begin{figure}
\begin{center}
\includegraphics*[width=8cm]{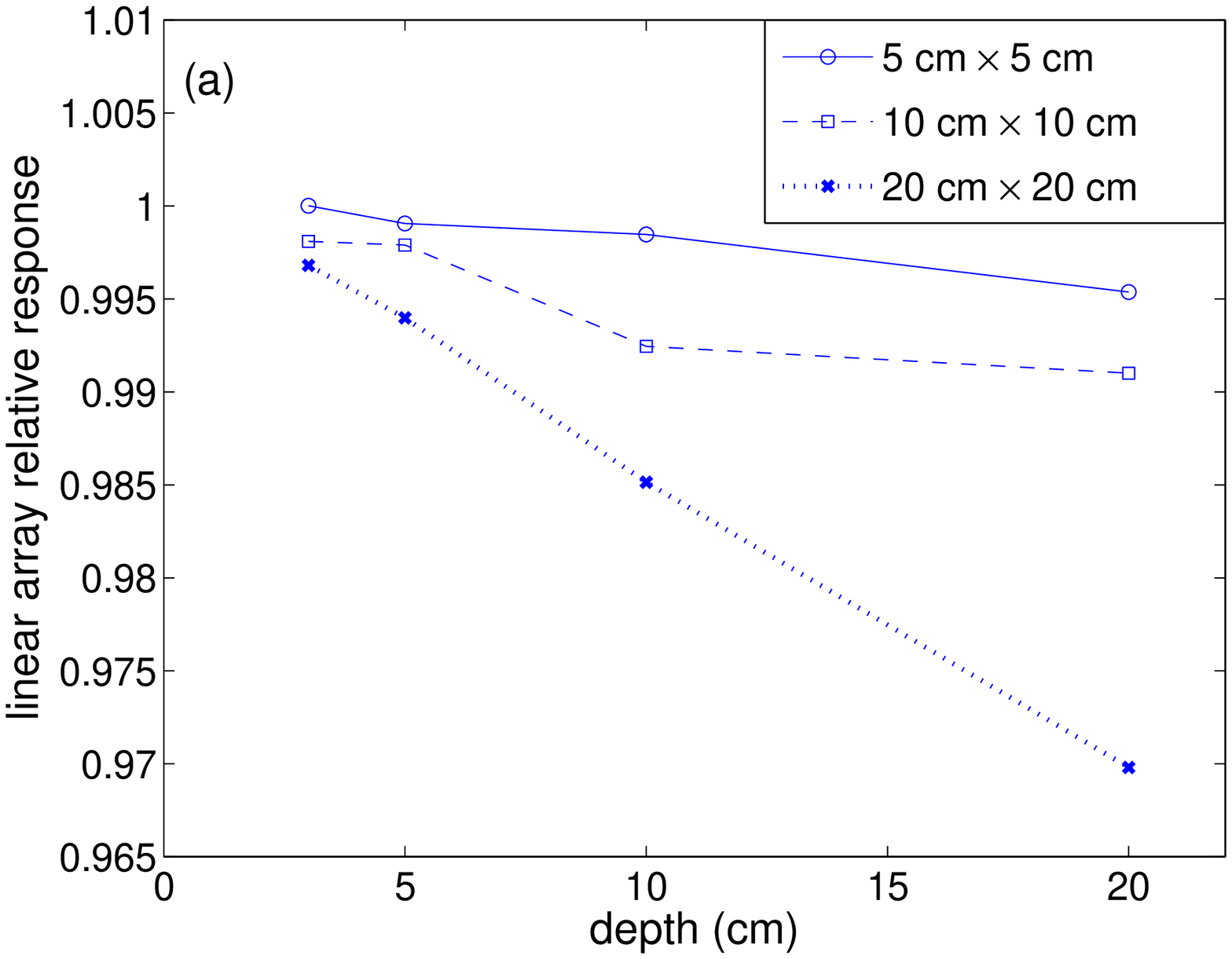}
\includegraphics*[width=8cm]{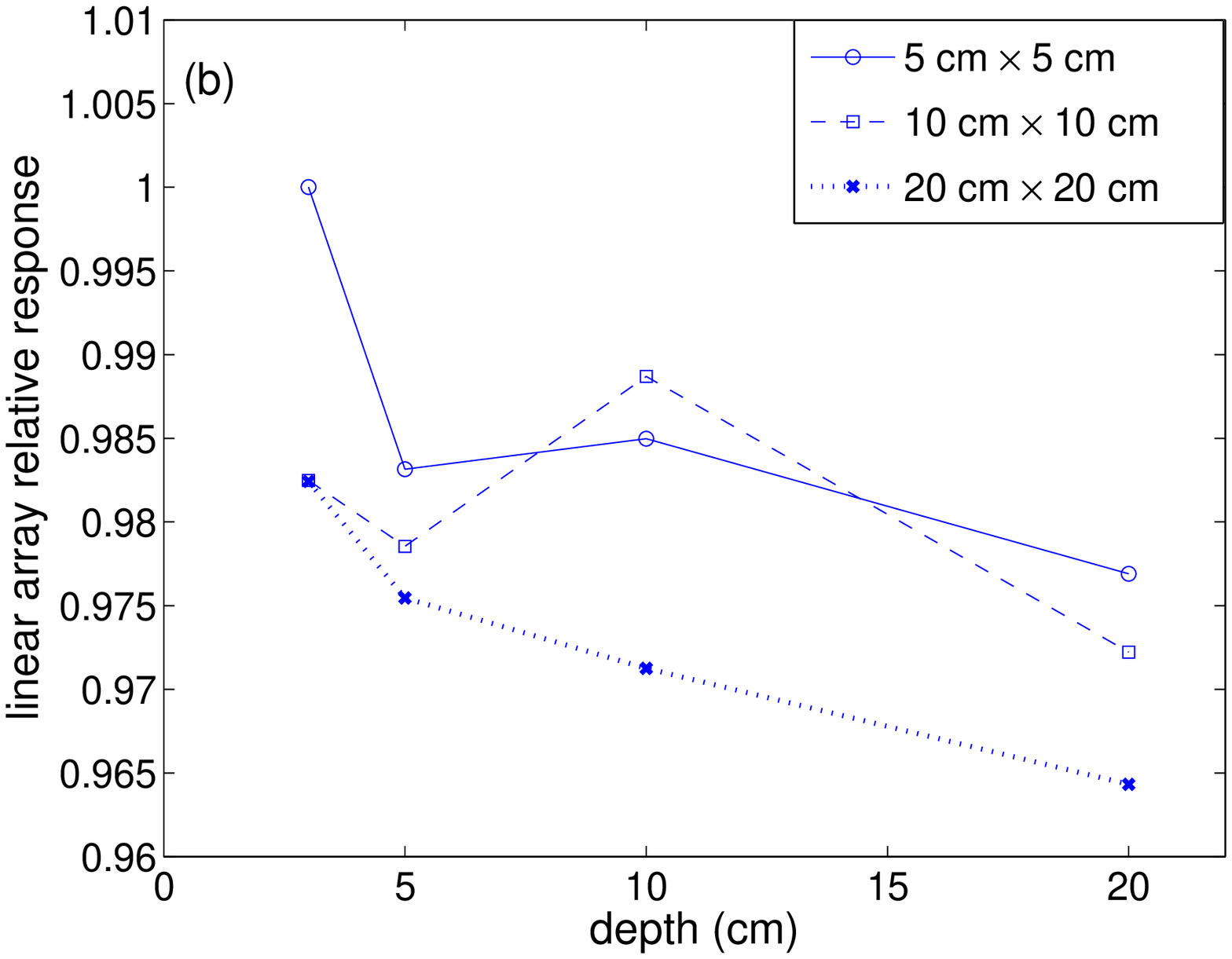}
\end{center}
\caption{Field size and depth dependence of the linear array response for 6 MV (a) and 15 MV (b).}
\label{fig10}
\end{figure}

To study the dependence on the energy spectrum of the incident radiation and the influence of the irradiated area on the read-out signal, 
measurements were performed for 6 MV and 15 MV at several depths and for several field dimensions. The results were compared with the
 data obtained with the 0.125 cm$^{3}$ chamber. Figure \ref{fig10} shows the ratio of the linear array measurements to those
 of the reference detector. The ratio was normalized to unit for a 5 cm $\times$ 5 cm field at 3 cm both for 6 MV and 15 MV. The relative uncertainties of
 the normalized data plotted in the figure are around 0.5 \%.
From this figure it is clear that the relative signal decreases when the field size increases as was expected from the OFs measurements. In addition the linear array
 underrresponds when the depth is increased. This under-response is up to 2.7 \% for 6MV and up to 2.5 \% for 15 MV, and again can be related with the metallization
 of the electrodes (Martens \etal 2001).

\subsection{Measurement of an IMRT and virtual wedge profiles}
\begin{figure}
\begin{center}
\includegraphics*[width=10cm]{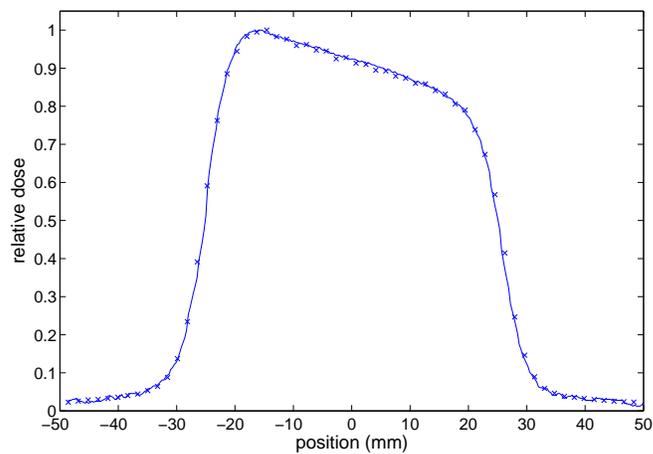}
\end{center}
\caption{45$^{0}$ virtual wedge relative dose profile of a 5 cm $\times$ 5 cm 6 MV photon beam delivering a total of 200 MU measured with the linear
 array ($\times$) and with RGF (solid line).}
\label{fig11}
\end{figure}

The detector can be used for the verification of virtual wedges. A profile of a 45$^{0}$ virtual wedge for a 15 MV 10 cm $\times$ 10 cm, delivering a total
 of 200 MU, has been acquired with the linear array and RGF. The results are compared in figure \ref{fig11}, with a maximum difference of 3\% and an average
 difference less than 1 \%.

\begin{figure}
\begin{center}
\includegraphics*[width=10cm]{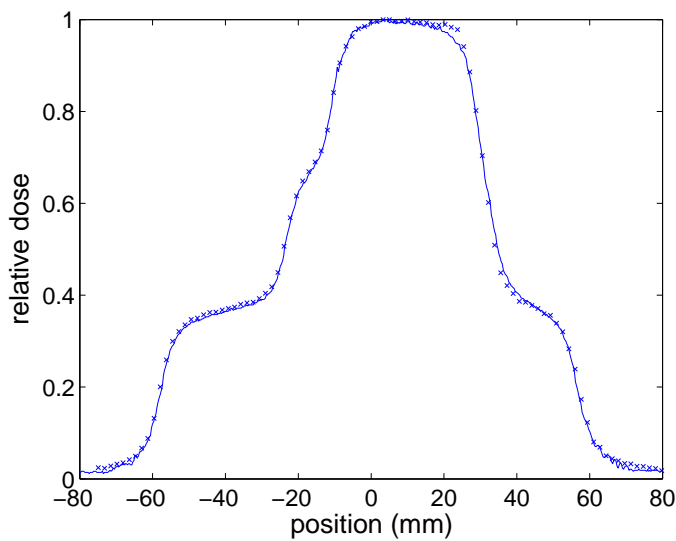}
\end{center}
\caption{Profiles for an IMRT field delivering a total of 200 MU measured with the linear array ($\times$) and with RGF (solid line).}
\label{fig12}
\end{figure}

Also a profile of an IMRT field which consists of four segments, each one delivering 50 MU, was acquired. Each segment delivered 50 MU. The linear array
 and RGF results are compared in figure \ref{fig12}. The maximum difference is 3 \% while the average difference is again less than 1 \%.

\subsection{Signal reproducibility}
The signal reproducibility was studied along the test period (around three months). All equivalent measurements were within a 2 \%. An important fraction of this deviation 
is due to temperature dependence, which has not been studied. Liquid-filled devices relative read-out signal dependence on temperature is around 10$^{-3}$ per degree due
 to the temperature influence
on  initial recombination (Mozumder 1974, Wickman \etal 1998).

\section{Conclusions}
The response of each linear array pixel is very linear with the dose rate (2.1 \% deviation at 5 Gy $min^{-1}$). A correction factor has to be applied to each pixel due to the low
 inhomogeneity in the XDAS response, and to small inhomogeneities in the gap and the pixel area.
OFs measurements deviates less than 1 \% from those measured with the reference detector for field widths from 2 cm to 10 cm (2 cm to 10 cm)
for 6 MV (15 MV). For the narrower fields the deviation is less than to 3 \%, but for this narrow fields the positioning uncertainty is high, and the difference
 between the active volume of a linear array pixel and the reference detector can affect the OFs measurements. 
The energy dependence is lower than 2 \% (for depths up to 20 cm and field sizes from 5 cm $\times$ 5 cm to 20 cm $\times$ 20 cm). 
Despite this dependence is not very large, it has to be taken into account when using the detector at high depth.

The detector has measured with accuracy several beam profiles and penumbras, and also IMRT and virtual wedge treatments. 
The small pixel size of the device combined with the fast and sensible XDAS read-out system allow a faster verification of these fields
 with a very good spatial resolution (even in regions of high dose gradient) and signal to noise ratio, making mechanical displacement unnecessary and
 showing its utility for high-precision relative dose measurements.

 In addition, the detector can be used for absolute dose measurements. The $G_{\rm{fi}}$ and its dependence with the electric field have
 been studied together with the charge losses due to volume recombination. Considering these effects, the absolute dose can be obtained from the read-out signal.
 Studies concerning the temperature dependence, the influence of the detector walls in the absolute dose deposited
 in the medium, the dose calibration and also the long term stability of the device will be the scope of further work.

\ack
This work has been supported by the research projects Xunta de Galicia PGIDT01INN20601PR and MCYT DPI2002-0185, and by a
 CIXTEC (Xunta de Galicia) grant.

\References

\item[] Belletti S, Cirio R, Cocuzza L, Degiorgis P G, Donetti M, Madon E, Marchetto F, Marletti M, Marzoli L, Peroni C, Trevisiol E and Urgesi A 2001 Pixel segmented ionization chamber for therapeutical beams of photons and hadrons {\it Nucl. Instrum. Methods A} {\bf 461} 420-1

\item[] Boag J W 1950 Ionization measurements at very high intensities: 1. Pulsed radiation beams {\it Br. J. Radiol.} {\bf 23} 601-11

\item[] Debye P 1942 Reaction rates in ionic solutions {\it Trans. Electrochem. Soc.} {\bf 82} 265-72

\item[] Eberle K, Engler J, Hartmann G, Hofmann R and H\"{o}randel J R 2003 First tests of a liquid ionization chamber to monitor intensity modulated radiation beams {\it Phys. Med. Biol.} {\bf 48} 3555-64
                                                                               

\item[] Fowler J F and Attix F H 1966 Solid state integrating dosimeters {\it Radiation Dosimetry} vol 2 (New York: Academic) 241-90

\item[] Johansson B, Wickman G and Bahar-Gogani J 1997 General collection efficiency for liquid isooctane and tetramethylsilane in pulsed radiation {\it Phys. Med. Biol.} {\bf 42} 1929-38

\item[] Jursinic P A and Nelms B E 2003 A 2-D diode array and analysis software for verification of intensity modulated radiation therapy delivery {\it Med. Phys.} {\bf 30} 870-9

\item[] Martens C, De Wagner C and De Neve W 2001 The value of the LA48 linear ion chamber array for characterization of intensity-modulated beams {\it Phys. Med. Biol.} {\bf 46} 1131-48

\item[] Martens C, Claeys I, De Wagner C and De Neve W 2002 The value of radiographic film for the characterization of intensity-modulated beams {\it Phys. Med. Biol.} {\bf 47} 2221-34

\item[] Mozumder A 1974 Effect of an external electric field on the yield of free ions. I General Results from the Onsager theory {\it J. Chem. Phys.} {\bf 60} 4300-4

\item[] Niroomand-Rad A, Blackwell C R, Coursey B M, Gall K P, Galvin J M, McLaughlin W L, Meigooni A S, Nath R, Rodgers J E and Soares C G  1998 Radiochromic film dosimetry: Recommendations of AAPM Radiation Therapy Committee Task Group 55 {\it Med. Phys.} {\bf 25} 2093-2115

\item[] Onsager L 1938 Initial recombination of ions {\it Phys. Rev.} {\bf 54} 554-7

\item[] Pardo J, Franco L, G\'{o}mez F, Iglesias A, Lobato R, Mosquera J, Pazos A, Pena J, Pombar M, Rodr\'{\i}guez A and Send\'{o}n J 2004 Free ion yield observed in liquid isooctane irradiated by $\gamma$ rays. Comparison with the Onsager theory {\it Phys. Med. Biol.} {\bf 49} 1905-14

\item[] Sykes J R, James H V and Williams P C 1999 How much does film sensitivity increase at depth for larger field sizes? {\it Med. Phys.} {\bf 26} 329-30

\item[] van Herk M and Meertens H 1988 A matrix ionization chamber imaging device for on-line patient setup verification during radiotherapy {\it Radiother. Oncol.} {\bf 11} 369-78

\item[] Wickman G and  Nystr\"{o}m H 1992 The use of liquids in ionization chambers for high precision radiotherapy dosimetry {\it Phys. Med. Biol.} {\bf 37} 1789-812

\item[] Wickman G, Johansson B, Bahar-Gogani J, Holmstr\"{o}m T and Grindborg J E 1998  Liquid ionization chambers for absorbed dose measurements in water at low dose rates and intermediate photon energies {\it Med. Phys.} {\bf 25}  900-7

\endrefs
\end{document}